\def\SiO      {{SiO$_2$ \/}}

\def\celcius  {{$^{\rm\circ}{\rm C}$\/}}

\newcommand{\ee}[1]{\cdot10^{#1}}

\newcommand{\unit}[1]{\,\mathrm{#1}}
\newcommand{\um}{\,\mu{\rm m}}
\newcommand{\us}{\,\mu{\rm s}}
\newcommand{\kT}{k_{\rm B}T}

\newcommand{\rtHz}{\sqrt{\rm Hz}}

\newcommand{\kc}{k_c}
\newcommand{\wc}{\omega_c}

\newcommand{\Fmin}{F_{\rm min}}

\newcommand{\captionstyle}{\normalfont} 

\documentclass[preprint,prl,aps]{revtex4}
\usepackage{graphicx}
\usepackage{bm}
\usepackage{comment}
\usepackage{amsmath}
\usepackage{amssymb}
\usepackage{verbatim}
\usepackage[colorlinks=true, pdfstartview=FitV, linkcolor=blue, citecolor=blue, urlcolor=blue]{hyperref} 
\usepackage{fmtcount}
\begin{document}



\title{Single-Crystal Diamond Nanomechanical Resonators with \linebreak Quality Factors $>1$ Million}

\author{Y. Tao$^{1,2}$, J. M. Boss$^{1}$, B. A. Moores$^{1}$, and C. L. Degen$^{1}$}
  \email{degenc@ethz.ch}
  \affiliation{
   $^1$Department of Physics, ETH Zurich, Schafmattstrasse 16, 8093 Zurich, Switzerland.
   $^2$Department of Chemistry, Massachusetts Institute of Technology, 77 Massachusetts Avenue, Cambridge MA 02139, USA.
   }
\date{\today}



\begin{abstract}
{\bf Diamond has gained a reputation as a uniquely versatile material, yet one that is intricate to grow and process \cite{balmer09}.  Resonating nanostructures made of single-crystal diamond are expected to possess excellent mechanical properties, including high quality factors and low dissipation \cite{sekaric02}.  Here, we demonstrate batch fabrication and mechanical measurements of single-crystal diamond cantilevers with thickness down to 85 nm, thickness uniformity better than 20 nm, and lateral dimensions up to 240 $\mu$m.  Quality factors exceeding one million are found at room temperature, surpassing those of state-of-the-art single-crystal silicon cantilevers of similar dimensions by roughly an order of magnitude \cite{yasumura00,mamin01}.  Force sensitivities of a few hundred zeptonewtons ($10^{-21}$N) result for the best cantilevers at millikelvin temperatures.  Single-crystal diamond could thus directly improve existing force and mass sensors by a simple substitution of resonator material.  Presented methods are easily adapted for fabrication of nanoelectromechanical systems (NEMS), optomechanical resonators, or nanophotonic devices that may lead to new applications in classical and quantum science.}
\end{abstract}

\maketitle

Nanomechanical resonators have led to pioneering advances in ultrasensitive sensing and precision measurements.  Applications range from nanoscale detection of forces in the context of scanning probe microscopy \cite{binnig86} to molecular-recognition-based mass screening in the medical sciences \cite{fritz00}.  The sensitivity of submicron-thick devices has progressed to a point where the attonewton force of single spins \cite{rugar04} or the mass of single molecules and proteins \cite{chaste12,hanay12} can be measured in real time.  Optomechanical resonators are furthermore intensively explored as quantum mechanical device elements in quantum science and technology \cite{poot12}.

A key figure of merit for a sensitive mechanical resonator is the rate at which it looses mechanical energy, described by the mechanical quality factor $Q$.  In the quantum mechanical regime, for example, the life time of an oscillator's vibrational ground state is $\tau = Q/\wc$, where $\wc$ the is resonator frequency.  The minimum detectable force (per unit bandwidth) is given by
\begin{equation}
\Fmin = \sqrt{\frac{4\kT\kc}{\wc Q}} ,
\label{eq:fmin}
\end{equation}
where $\kc$ is the spring constant of the oscillator and $\kT$ is thermal energy.  State-of-the-art silicon cantilevers achieve force sensitivities of typically $10^{-16}\unit{N/\rtHz}$ at room temperature and $10^{-18}\unit{N/\rtHz}$ at millikelvin temperatures \cite{mamin01}.  These force sensitivities are limited by thermomechanical noise in the resonator that arises from the dissipative coupling of the mechanical mode to the heat bath.

Despite considerable effort, attempts to improve sensitivities of resonators well into the zeptonewton-range $(1\unit{zN}=10^{-21}\unit{N})$ have not been particularly successful.  One strategy has been the development of thinner, more compliant resonators with lower spring constant $\kc$ ($\mu$N/m--mN/m).  The projected gain in sensitivity, however, was found to be counteracted by a decrease in the mechanical $Q$ with decreasing thickness due to surface friction \cite{yasumura00,yang02}.  Consequently, the best demonstrated force sensitivities have only marginally improved over the last decade \cite{mamin01,usenko11}.  Several other routes have been explored, including surface cleaning under UHV conditions \cite{rast06}, and use of alternative structures like doubly-clamped beams at high spring tension \cite{teufel09} or suspended carbon nanotube oscillators \cite{chaste12}.  While these efforts have produced exciting results, they either have not yet led to an improvement in force sensitivity, or are ill-suited for force sensing due to high non-linearities, incongruous device geometry, or unfavorable operating conditions.

Another strategy, followed here, is the fabrication of resonators from materials with very low intrinsic dissipation.  A particularly interesting choice is  single-crystal diamond:  with its remarkable mechanical strength, chemical resistance, optical transparency, and thermal conductivity, diamond has gained privileged niches in industrial and engineering applications.  Most of these properties are ultimately rooted in the extreme strength and stability of the $sp^3$ carbon network of the diamond crystalline lattice.  Moreover, diamond does not form any surface passivation layer (as does silicon) apart from a monolayer of covalently bound terminating surface atoms.  Unfortunately, fabrication of diamond into single-crystalline NEMS has been hampered by the difficult growth of the material, its resistance to processing, and only over the last few years has material amenable to nanofabrication become available \cite{balmer09}.

To illustrate the challenge in fabricating single-crystal diamond nanostructures, one may note that the material cannot be grown on any other substrate than single-crystal diamond itself \cite{balmer09}.  This property precludes wafer-scale processing.  Early diamond NEMS have therefore focused on polycrystalline diamond that can be readily grown as a thin film on various substrates, including silicon.  In 2004, researchers demonstrated fabrication of 40 nm-thin devices (grain size 10-100 nm) exhibiting Q-factors up to 10,000 at cryogenic temperatures \cite{hutchinson04}.
Only recently have researchers tackled fabrication of single-crystal diamond MEMS and NEMS.  In the absence of wafer material, the main route to producing thin films has relied on ion bombardement.  In this approach, a flat diamond surface is exposed to a large dose of high energy (MeV) ion irradiation, resulting in a damage layer typically a few hundred nanometers below the surface that can be selectively etched to peel off thin membranes \cite{parikh92}.  Impinging ions, however, have to travel through the device layer, leaving a large number of defects in their track, thus degrading the material.  While the issue might be circumvented by re-growth of pristine material \cite{aharonovich12}, quality factors of ion-irradiation-fabricated membranes have so far been limited to $Q\lesssim20,000$ \cite{zalalutdinov11}.

In this contribution, we report on two significant advances made towards ultrasensitive diamond nanomechanical resonators.  In a first part, we present two strategies to fabricate high-quality and high-aspect ratio nanocantilevers from ultrapure, single-crystalline diamond starting material, achieving a processing level comparable to silicon.  In a second part we show that these procedures lead to nanomechanical resonators with exceptional quality factors and very low intrinsic dissipation.  Finally, we gain significant insight into the underlying physical dissipation mechanisms that provide rational basis and strategies for further device improvement.

\begin{figure}[h!]
\centering
\includegraphics[width=0.98\textwidth]{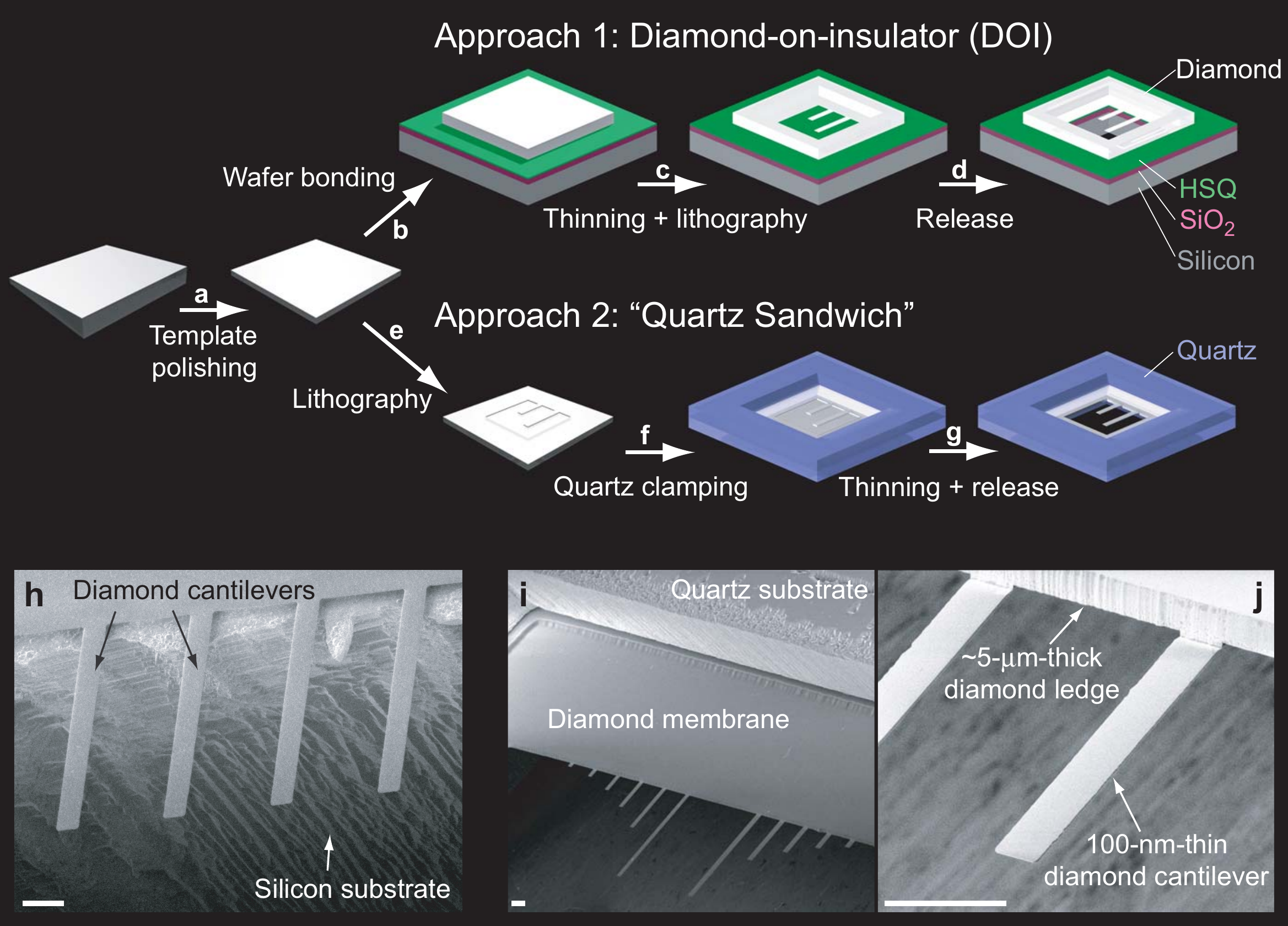}
\caption{\captionstyle
{\bf Batch Fabrication of single-crystal diamond nanocantilevers.}
{\bf a}. A roughly $3\times3\times0.03\unit{mm^3}$ single-crystal diamond plate is template-polished to a thickness uniformity $<1\um$ over the entire plate.
{\bf b}. {\bf First approach:} Plate is wafer-bonded using hydrogen silsesquioxane (HSQ) to form a diamond-on-insulator (DOI) substrate.
{\bf c}. Plate is thinned to $0.1-1\unit{\um}$ thickness by reactive ion etching, and cantilevers are patterned using optical lithography.
{\bf d}. Cantilevers are released using conventional back-side etching.
{\bf e}. {\bf Second approach:} Cantilevers are patterned using optical lithography.
{\bf f}. Diamond plate is clamped between two fused quartz slides with a $\sim2\times2\unit{mm^2}$ central aperture.
{\bf g}. Exposed diamond plate is etched down until cantilevers become released.
{\bf h}. Scanning electron micrograph of finished DOI devices.
{\bf i,j}. Scanning electron micrographs of finished ``quartz sandwich'' devices.  Scale bars are $20\unit{\um}$.
}
\label{fig:fabrication}
\end{figure}
The basic fabrication pathway is shown in Figure \ref{fig:fabrication}.  Our route taken starts with mm-sized single-crystal diamond plates of $20-40\um$ thickness and $<100>$ surface orientation grown by chemical vapor deposition \cite{balmer09}.  The advantage of these plates is the high material quality (single crystal, low doping) and the fact that they are commercially available.  The large thickness variation (up to $10\um$ over the entire plate) and the delicate handling, however, are significant obstacles if large-area, sub-micron-structures are to be made.  We have therefore developed a template-assisted re-polishing procedure to improve thickness uniformity to $<1\um$ based on a polycrystalline diamond mold \cite{supplementary}.  To facilitate handling, we have implemented two different bonding strategies: In a first approach, we achieved direct wafer bonding to a thermal oxide-bearing silicon substrate using hydrogen silsesquioxane (HSQ) resist as intermediary.  The advantage of this approach is a generic diamond-on-insulator (DOI) substrate amenable to any type of follow-up lithography.  In a second strategy, we have clamped diamond plates between two \SiO substrates resulting in a fused ``quartz sandwich'' structure.  The advantage of this method is the simpler and faster processing exploiting quartz as both the handling substrate and mask material.  In a next step, the diamond device layer was thinned to \textless 500 nm using reactive ion etching based on an argon-chlorine plasma \cite{lee08}.  This procedure is known to produce very uniform and smooth etches with resulting surface roughness well $<$1 nm-rms.  Cantilevers were then defined by standard optical lithography before they were fully released by backside etching steps.  For samples made with the quartz sandwich method, a thick diamond ledge ($\sim5\unit{\um}$) was conserved at the base of the cantilevers to reduce clamping losses, while for DOI devices, the SiO$_2$ layer served as the clamping structure.
Micrographs of several devices are shown in Figure \ref{fig:fabrication}(h-j).  Further details of the fabrication procedure are given as supplementary information \cite{supplementary}.

For the present study we have fabricated four single-crystal diamond chips bearing roughly $\sim25$ cantilevers each.  Three devices were made using the quartz sandwich method and one using the DOI method.  We found both methods to be highly robust, achieving an overall 106 out 108 cantilever yield.  Of the four devices, three were fabricated from ``optical-grade'' starting material (Delaware Diamond Knives Inc.) with a doping concentration $[N]<1\unit{ppm}$, and a fourth chip was made from ``electronic-grade'' material (ElementSix) with a much lower doping of $[N]<5\unit{ppb},[B]<1\unit{ppb}$ \cite{supplementary}.  Cantilevers were between 20 and 240$\um$ long, between 8 and 16$\um$ wide, and between 80 and 800 nm thick.  Thickness variation along the entire length was found to be less than 100 nm, even for the longest 240-$\um$-cantilevers.  Corresponding resonance frequencies were between $\wc/(2\pi) = 2\unit{kHz}-6\unit{MHz}$ and spring constants were between $\kc=60\unit{\mu N/m}-300\unit{N/m}$.  Since the lowest dissipation levels result for long and thin (high aspect ratio) structures we focused on those resonators. 

The large number of finished devices allowed us to obtain a detailed picture of the mechanical dissipation in these structures.  In particular, we have assessed the dependence of the quality factor on geometry, surface termination, temperature, and doping concentration.  For these measurements we used a custom-built force microscope apparatus operated under high vacuum ($<10^{-6}\unit{mbar}$) mounted at the bottom of a dilution refrigerator ($80\unit{mK}-300\unit{K}$).  The high vacuum eliminated viscous (air) damping and the refrigerator both served for temperature-dependence studies and for minimizing thermomechanical noise.  Quality factors $Q$ were measured by the ring-down method \cite{yasumura00} using a low-power ($\leq 10\unit{nW}$) fiber-optic interferometer for motion detection \cite{mamin01}.  More details on the experimental setup are provided in the Methods section.

\begin{figure}[h!]
\centering
\includegraphics[width=0.98\textwidth]{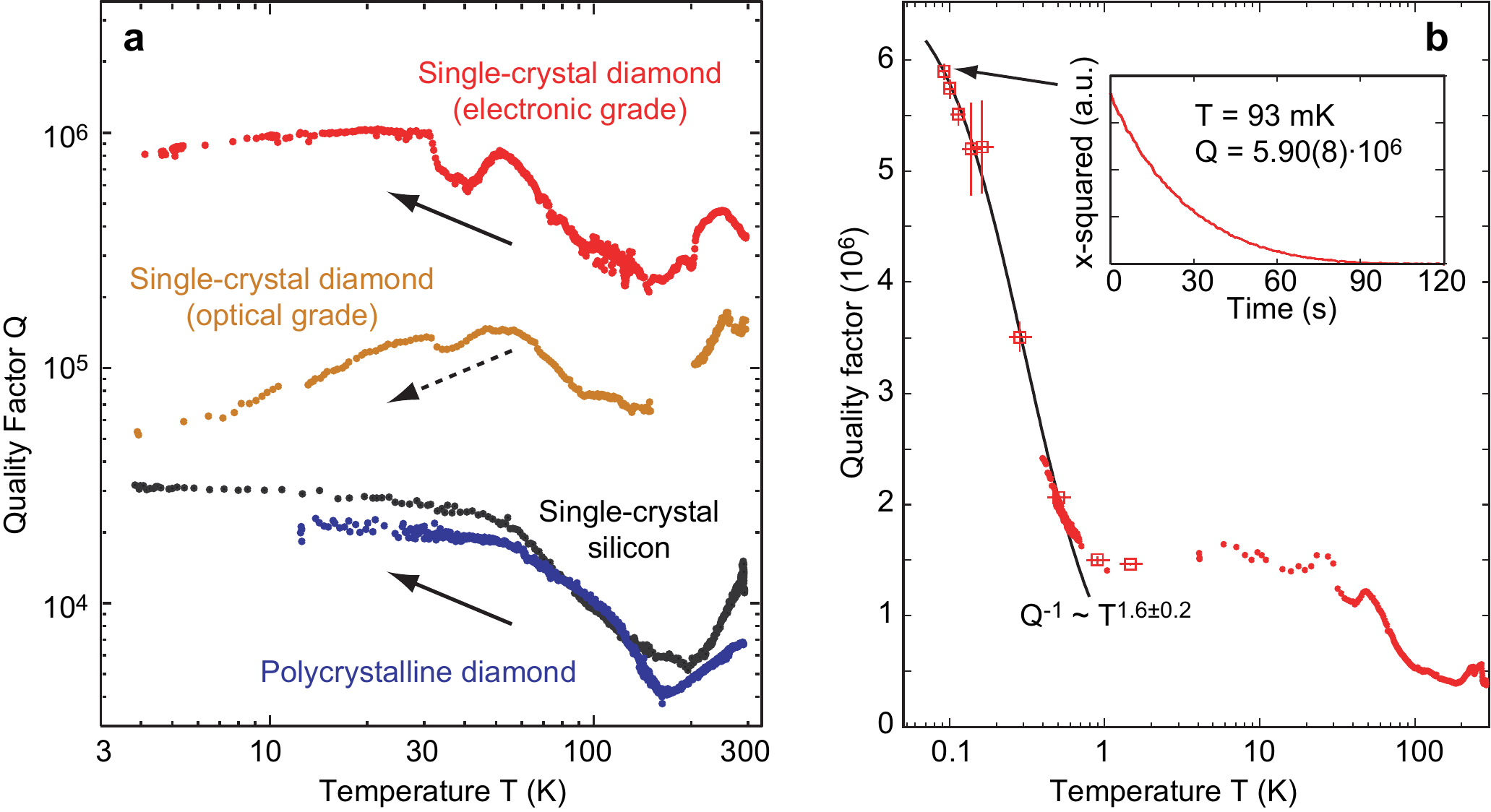}
\caption{\captionstyle
{\bf Quality factors of single-crystal diamond nanoresonators between 0.1-300 K.}
{\bf a.} Two representative diamond devices (out of 9 measured) are compared to refererence devices made from polycrystalline diamond and single-crystal silicon of similar thickness ($100-300\unit{nm}$).  $Q$ factors between 100'000-1'000'000 are observed for diamond devices at room temperature, roughly 10-100$\times$ higher than the reference devices.  Cooling to 3 K leads to an increase in $Q$ for the electronic grade (low doping) as well as the two reference resonators that are dominated by surface friction (solid arrows).  Conversely, a reduction is seen in $Q$ for the optical-grade (high doping) diamond resonator that has a strong contribution of bulk friction (dashed arrow).
{\bf b.} Quality factor of a 660-nm-thick electronic-grade resonator in the millikelvin regime.  Red dot data are obtained by sweeping refrigerator temperature, and red square data (with error bars) are obtained by varying the laser power incident at the resonator (see Methods).  Solid black line is a power-law fit with $Q^{-1}\propto T^{1.6\pm0.2}$.  Inset shows the ring-down measurement at $T=93\unit{mK}$, yielding a Q factor of $5.9$ million and a corresponding force sensitivity of about $540\unit{zN/\rtHz}$.
Additional parameters for all cantilevers can be found in Table \ref{table:cantilevers}.
}
\label{fig:temperature}
\end{figure}

In a first set of experiments we have measured the temperature-dependent quality factors between 3-300 K of 9 diamond resonators total.  Two representative measurements, one of an electronic grade (low doping) and one of an optical grade (high doping) device, are shown in Figure \ref{fig:temperature}a.  A third and fourth curve of a polycrystalline diamond cantilever and an ultrasensitive single-crystal silicon cantilever are added for reference.  The polycrystalline cantilever was produced from 3-5 nm grain size material (Advanced Diamond Technology) \cite{supplementary}, while the silicon cantilever was identical to those used in single-spin detection experiments \cite{mamin01,rugar04} and served as a ''best-of-its-kind'' benchmark device.  Although only two diamond curves are shown, we found the general features visible in Figure \ref{fig:temperature}a to reproduce well between replicates and fabrication methods (see supplementary information).

Figure \ref{fig:temperature}a immediately reveals several interesting features.  First, and most strikingly, we observe that single-crystal diamond resonators show very high quality factors.  For instance, the room-temperature $Q$ values of the two diamond resonators shown are 150,000 and 380,000 with a device thickness of only 100 nm and 280 nm, respectively (see Table \ref{table:cantilevers}.  Other resonators showed room-temperature $Q$ factors up to 1.2 million at a thickness of 800 nm (see below).  These values are between one and two orders of magnitude higher than similar polycrystalline \cite{hutchinson04} and single-crystal silicon resonators \cite{yasumura00,yang02}.

Second, Figure \ref{fig:temperature}a shows a clear difference between the electronic-grade (low-doping) and optical-grade (high-doping) single-crystal diamond resonators.  Most prominently, we observe that the $Q$ factor of the electronic-grade resonator (and the two reference devices) {\sl decreases} towards cryogenic temperatures while that of the optical-grade resonator {\sl increases}.  Our understanding of this difference is as follows (more evidence will be given below):  at room temperature, mechanical dissipation of all resonators is limited by surface friction.  Surface friction is the most common dissipation mechanism for sub-micron-thick cantilevers \cite{yasumura00,yang02} and commonly attributed to surface passivation layers or adsorbate molecules.  Since all devices show a similar behavior as they are cooled from 300 K, the surface dissipation mechanism appears generic, such as related to common adsorbates.  As temperatures approach 3 K, surface friction is markedly reduced (reflected in higher $Q$ values) but it remains the dominating dissipation mechanism.  The only exception is the optical-grade resonator:  Here, a second friction mechanism begins to dominate at 3 K, reflected in lower $Q$ values.  Since the only nominal difference between the electronic- and optical-grade diamond resonators is their doping concentration, it would be natural to assume that bulk impurities are related to the increase in friction.  

To explore the potential of diamond nanomechanical resonators for millikelvin applications \cite{rugar04,poot12}, we have investigated dissipation in another electronic-grade resonator down to about $0.1\unit{K}$.  These data are shown in Figure \ref{fig:temperature}b.  We find that below about one kelvin, the $Q$ factor further increases and for this particular device attains a value of almost six million at base temperature ($\sim93\unit{mK}$).  This is the highest $Q$ factor we have observed in this study, and to the best of our knowledge, the highest $Q$ factor reported for any sub-micron-thick cantilever resonator to date.  Since $Q$ is still increasing towards the lowest temperatures, there is scope that even higher values could result in the ten-millikelvin regime.
\begin{table*}[t!]
\centering
{\footnotesize
\begin{tabular}{|c|c c c|c c|c c|c c|}
\hline
\hline
Material$^a$	& Length  & Width   & Thickness$^b$   & $\wc/2\pi$ 	& $\kc$       & $Q$ (300 K) & $Q$ (3 K) & $\Fmin$ (300K) & $\Fmin$ (3K) 	\\
        			& ($\um$) & ($\um$) & (nm)      			& (Hz)			  & (mN/m)      &             &           & (aN/$\rtHz$)   & (aN/$\rtHz$)   \\
\hline
el-SCD 	  		& 240     & 12      & 280(20)  				&  13,168 		& 4.8 				& 380,000     & 800,000   &  50            & 3.5	  				\\
o-SCD			   	& 120 		& 12      & 100(25)  			 	&  15,097  		& 1.4         & 150,000     & 55,100    & 40             & 6.6 						\\
Si		  		  & 170 		& 4       & 135(5)  				&   4,960  		& 0.083       &  11,500     & 31,500    & 63             & 3.8    				\\
PCD 		  		& 200 		& 18      & 270(5)   				&  14,197  		& 5.0      		&  6,680      & 22,000    & 370            & 21   					\\
\hline
el-SCD		 		& 240     & 12      & 660(20)   			&  32,140 		& 67      		& 412,000     & 1,510,000 & 115            & 6.0 (0.54$^c$)	\\ \hline
\hline
\end{tabular}
}
\caption{\captionstyle
{\bf  Mechanical properties for selected cantilevers.}
$^a$el-SCD = electronic-grade single-crystal diamond, o-SCD = optical-grade single-crystal diamond, Si = single crystal silicon, PCD = polycrystalline diamond.
$^b$Thickness values are reported as the average between the thickest and thinnest parts of each cantilever along its length, with values in parentheses denoting the difference between this average and the extrema.  For most diamond devices, the extrema occurred at the base and the tip.
$^c$At 93 mK.  
}
\label{table:cantilevers}
\end{table*}
%

\begin{figure}[h!]
\centering
\includegraphics[width=0.95\textwidth]{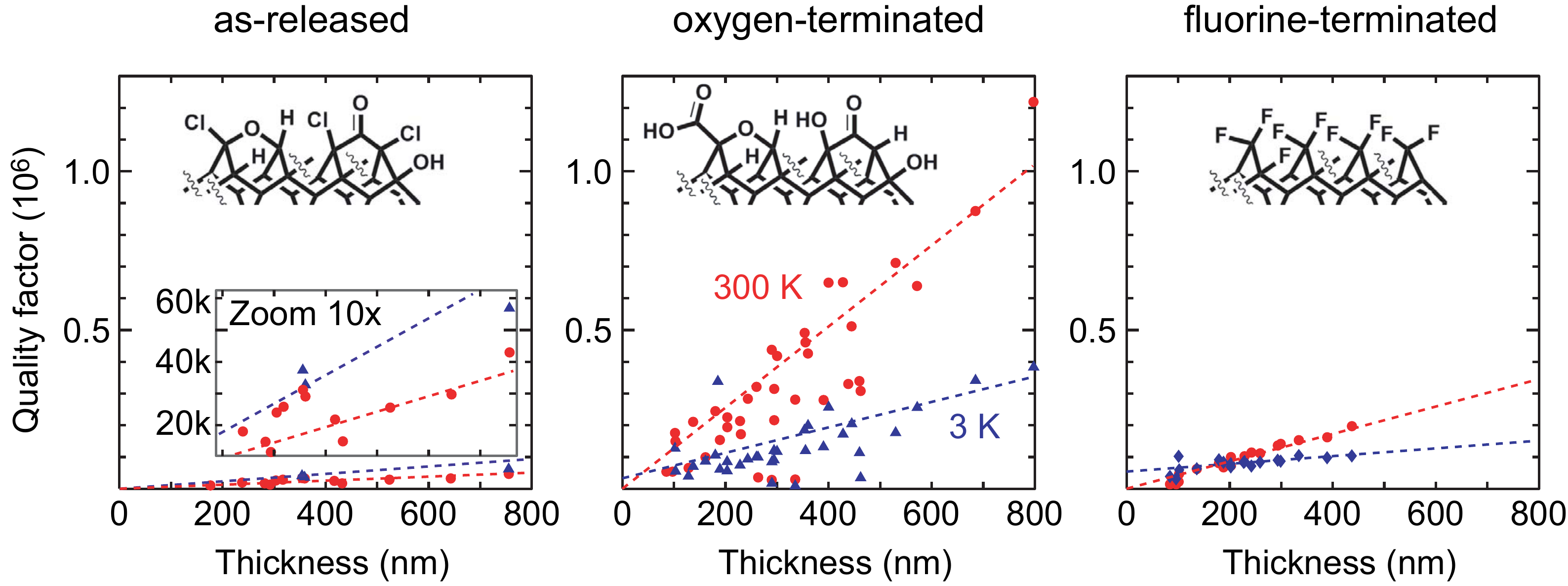}
\caption{\captionstyle
{\bf Impact of surface chemistry on dissipation.}
Three different chemical surface terminations (sketched) are investigated on 15-40 devices of an optical-grade chip.  Red dots are 300 K values and blue triangle are 3 K values, and dashed lines serve as guides to the eye.  Large variation of $Q$ factors is seen as surface chemistry and thickness are changed, underlining the critical role of surface friction.  Very high $Q$ factors, exceeding one million at room temperature, are found for oxygen-terminated devices.  
}
\label{fig:surface}
\end{figure}

To corroborate these findings and to further investigate the dissipation mechanism we have surveyed between 15 and 40 resonators under three different chemical surface terminations.   This second set of measurements served to explore whether surface chemistry affects dissipation, and to find out whether $Q$ factors can be improved by proper choice of surface termination.  In a first round, cantilevers were measured ``as-released'' right after the final argon-chlorine plasma reactive ion etch.  In this state, the surface has a mixture of covalently attached elements, including H, O and Cl (see atomistic sketch in Figure \ref{fig:surface}).  In a second round, the surface was oxygen-terminated using low-temperature (450 \celcius) annealing in air \cite{supplementary}.  O-termination represents the standard hydrophilic surface termination of diamond \cite{sque06}.  In a third round, the surface was fluorine-terminated using CF$_4$ plasma \cite{supplementary}.  F-termination is known to produce a simple monolayer coverage that is both hydrophobic and oleophobic with potentially very little molecular adsorption.  We found O-terminated cantilevers to be very stable with no measurable degradation in $Q$ after a two-months exposure to ambient atmosphere (F-terminated cantilevers were not measured).

The data of these measurements are shown in Figure \ref{fig:surface} and are grouped by surface termination.  We immediately see that the particular surface chemistry has a strong impact on dissipation, with a more than tenfold variation between ''as-released'' and O-terminated devices.  While the improvement with O- and F-termination compared to the ''as-released'' state can be understood in terms of cleaner surfaces with better-defined surface chemistries \cite{supplementary}, the superior performance of O-terminated compared to F-terminated devices poses more challenges to interpretation.  One hypothesis is that the terminating atoms are directly responsible for dissipation with little influence of adsorbates, and that $F$-atoms more efficiently engage in energy relaxation.  Another hypothesis is that the inductive withdrawing effect of the polar C-F chemical bonds stabilizes a layer of electronic defects right below the surface that enhances energy relaxation \cite{panich10}.  If the second interpretation were true, it would be interesting to examine less polar termination groups including chlorine, sulfur, amine, hydrogen, and alkyl groups \cite{miller99}.

Besides the variation with surface chemistry, we also observe that most plots in Figure \ref{fig:surface}(a-c) show a roughly linear relationship between $Q$ and thickness (shown by dashed lines).  A thickness dependence of $Q$ indicates a surface-related dissipation mechanism, whereas no thickness dependence would be characteristic of a bulk friction mechanism.  (We further observed a slight length dependence of $Q$, indicating that some support or clamping loss may be also present for shorter devices \cite{supplementary}).  The linear relationship between $Q$ and thickness is most pronounced at 300 K (red data); at 3 K (blue data), the linear dependence is much weaker and $Q$ values between O-terminated and F-terminated resonators converge.  These observations confirm the picture from the temperature-dependence study in Figure \ref{fig:temperature}a:  at room temperature, all resonators are limited by surface friction independent of surface chemistry, while at 3 K, bulk friction is strongly contributing and mostly dominating for F-terminated devices.

\begin{figure}[h!]
\centering
\includegraphics[width=0.45\textwidth]{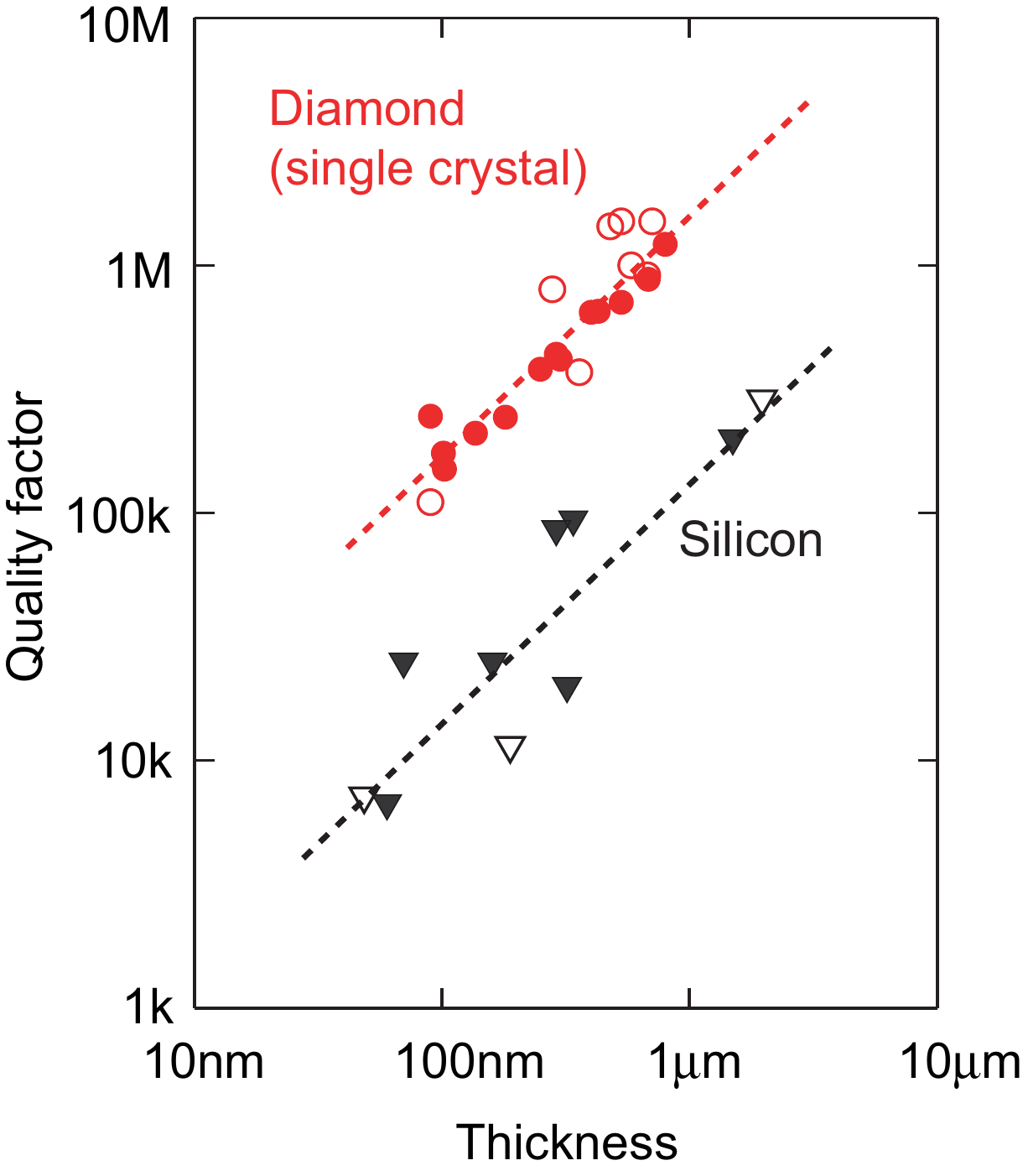}
\caption{\captionstyle
{\bf Comparison between diamond and silicon mechanical resonators,} highlighting that for similar device dimensions, quality factors are consistently higher by about an order of magnitude.
Open symbols are 300 K values and filled symbols are $\sim$ 4 K values.  Dashed lines indicate linear thickness dependence.  Diamond data are from this study, silicon 4-K data are taken from a compilation in Ref. \cite{yasumura00}, and silicon 300-K data are from our earlier measurements.  Data references are given in the supplementary information (Ref. \cite{supplementary}).
}
\label{fig:silicon}
\end{figure}

To exemplify the advances possible with diamond NEMS, we have compiled a number of $Q$ values from previous studies on ultrasensitive silicon resonators and plotted them alongside with the diamond data from Figure \ref{fig:surface}.  The two data sets are shown in Figure \ref{fig:silicon}.  The comparison highlights that diamond could offer a consistent order-of-magnitude improvement in mechanical $Q$ compared to single-crystal silicon without changing the geometry of a particular NEMS device.

It is instructive to evaluate the mechanical dissipation levels of diamond resonators in the light of applications in quantum nanomechanics or ultrasensitive force sensing.  For example, taking the electronic-grade resonator from Figure \ref{fig:temperature}b, we can calculate a mean quantum-mechanical occupation number $\bar{n} \approx \kT/\hbar\wc $ and a thermal decoherence time $\tau = \hbar Q/\kT$.  At $T=93\unit{mK}$ and $Q=5.9\ee{6}$, we obtain $\bar{n}\approx6\ee{4}$ and $\tau\approx0.5\unit{ms}$.  This decoherence time is already considerably longer than one oscillation period $2\pi/\wc\approx31\unit{\us}$, even if the resonator is still far from the quantum mechanical ground state, highlighting the potential of the material for quantum applications.  Numbers could be much improved by using feedback cooling \cite{poot12} or using motion transducers compatible with $\sim10\unit{mK}$ bath temperatures \cite{usenko11}.  The main advantage of the diamond resonators compared to other high-$Q$ resonators is their high compliance that will allow for much stronger couplings.  

We can also convert the measured quality factors into values of force sensitivity.  For the cantilever shown in Figure \ref{fig:temperature}b, we find $\Fmin = 115\unit{aN/\rtHz}$ at room temperature and $540\unit{zN/\rtHz}$ at 100 mK (see Table \ref{table:cantilevers} and Ref. \cite{supplementary}).  Other cantilevers in this study showed force sensitivities as low as $26\unit{aN/\rtHz}$ at room temperature and $3.5\unit{aN/\rtHz}$ at 3 K.  These force sensitivities are remarkable considering that the geometry of these test devices is not particularly optimized.  With the present diamond material and processing method, we are confident that a cantilever thickness as small as $50\unit{nm}$ and a width of $1\unit{\um}$ could be realized for a 240 $\um$-long cantilever.  Such a cantilever has a projected force sensitivity of $9.4\unit{aN/\rtHz}$ at 300 K, $490\unit{zN/\rtHz}$ at 3 K and $45\unit{zN/\rtHz}$ at $100\unit{mK}$, based on the scaling of $\Fmin$ with geometry \cite{supplementary} and the linear thickness dependence of $Q$.  To put this into perspective, if the latter sensitivity could be carried over to spin detection experiments \cite{rugar04,degen09}, it would be possible to detect a single electron spin in about 1 ms and a single proton spin in 1 min at unit signal-to-noise-ratio.

In conclusion, we have presented measurements of mechanical dissipation in sub-micron-thick single-crystal diamond nanomechanical resonators.  We find that diamond is an excellent NEMS material, with $Q$ factors and dissipation levels exceeding state-of-the-art silicon devices by roughly one order of magnitude.  Measurements of the $Q$ factor at room and cryogenic temperatures furthermore underline the importance of both surface quality and low-defect bulk material.  Several possible avenues for reducing dissipation exist in both departments, including different surface chemistries \cite{miller99}, atomic surface flatness \cite{watanabe99}, reduction of intrinsic defects through high-temperature annealing \cite{davies92}, or even the use of isotopically pure diamond material \cite{ishikawa12}.  Moreover, we expect that geometry can be significantly optimized without unduly compromising device yields, moving force sensitivities from the current $\sim 540\unit{zN/\rtHz}$ well into the $10-100\unit{zN/\rtHz}$-range.

We believe that the fabrication methods reported herein are of potentially very broad applicability.  The facile and robust wafer-bonding approaches presented result in general diamond-on-insulator (DOI) or quartz-bonded substrates that are amenable to essentially any type of follow-up lithography.  Due to its low absorption and high refractive index, for example, high-quality single-crystal diamond would be ideally suited for use in optomechanical cavities or as photonic crystal material \cite{burek12}.  Diamond further possesses interesting lattice defects, such as nitrogen-vacancy centers, that could be directly embedded in high-$Q$ diamond NEMS \cite{maletinsky12}.  Such on-chip devices are considered key components for hybrid quantum systems in quantum science and technology, underscoring the prominent material platform diamond can offer to future applications.


{\bf Methods:}
Mechanical properties of diamond nanoresonators were measured in a custom-built scanning force microscope designed for magnetic resonance force microscopy \cite{rugar04}.  Cantilevers were prepared under ambient conditions and then mounted in a high-vacuum chamber ($<10^{-6}\unit{mbar}$) at the bottom of a dilution refrigerator ($\sim80\unit{mK}-300\unit{K}$).  Resonator frequency $\wc$ and quality factor $Q$ were measured using the ring-down method \cite{yasumura00}, and the spring constant $\kc$ calibrated via a thermomechanical noise measurement at room temperature, and sometimes, at 3 K \cite{mamin01}.  As a consistency check $\wc$ and $\kc$ were independently calculated from the geometry using a finite element software (COMSOL).  Resonator motion was detected using a low-power fiber-optic interferometer  operating at a wavelength of 830 nm and producing less than $10\unit{nW}$ of laser light incident at the cantilever.  To exclude cavity effects, it was verified that the same $Q$ factor was obtained whether the measurement was done on the positive or negative (red- or blue-shifted) side of the interferometer fringe.
To measure temperature dependence of quality factors, two complementary measurement techniques were employed.  For temperatures between $0.4-300\unit{K}$, $Q(T)$ was measured the usual way by slowly ($\lesssim 0.2\unit{K/min}$) sweeping refrigerator temperature and assuming thermal equilibrium between resonator and bath.  For very low temperatures ($T<2\unit{K}$) the refrigerator was operated at base temperature (80 mK) and resonator temperature was adjusted by varying interferometer laser power through absorptive heating (see supplementary information).  The advantage of the latter method is that resonator temperature can be directly estimated via the low-temperature thermal conductivity of diamond. (Cantilever temperature could also be inferred from a thermomechanical noise measurement \cite{mamin01}, but this method was inaccurate in our experiments due to slight mechanical vibrations introduced by the dilution circuit.)

{\bf Acknowledgments:}
This work was supported by the NCCR QSIT, a competence center funded by the Swiss NSF, Swiss NSF grant 200021\_137520/1, and ERC Starting Grant 309301.  We thank Joseph Tabeling, Peter Morton and the technical staff at DDK for assistance with diamond polishing and ElementSix for providing diamond samples through the DARPA QuASAR program.  We thank the clean room staff at FIRST Lab and CLA (ETH), IBM Zurich, and the MTL at MIT for advice on fabrication.  We thank the ETH and MIT machine shops for help in construction of the measurement apparatus.  We thank Cecil Barengo, Kurt Broderick, Ute Drechsler, Ania Jayich, Gang Liu, Paolo Navaretti, Martino Poggio, Donat Scheiwiller, and Dave Webb for technical assistance and useful discussions.





\noindent
\begin{thebibliography}{99}


\bibitem{balmer09}
  Balmer, R. S. {\sl et al.}, Chemical vapor deposition synthetic diamond: Materials technology and applications.
  \href{http://dx.doi.org/10.1088/0953-8984/21/36/364221} {{\sl J. Phys. Cond. Matt.} {\bf 21}, 364221 (2009)}.
\bibitem{sekaric02}
  Sekaric, L. {\sl et al.}, Nanomechanical resonant structures in nanocrystalline diamond.
  \href{http://dx.doi.org/10.1063/1.1526941} {{\sl Appl. Phys. Lett.} {\bf 81}, 4455-4457 (2002)}.
\bibitem{yasumura00}
  Yasumura, K. Y. {\sl et al.}, Quality factors in micron- and submicron-thick cantilevers.
  \href{http://dx.doi.org/10.1109/84.825786} {{\sl J. Microelectromech. Syst.} {\bf 9}, 117 (2000)}.
\bibitem{mamin01}
  Mamin, H. J. \& Rugar, D., Sub-attonewton force detection at millikelvin temperatures.
  \href{http://dx.doi.org/10.1063/1.1418256} {{\sl Appl. Phys. Lett.} {\bf 79}, 3358 (2001)}.


\bibitem{binnig86}
  G. Binnig, C. F. Quate, and C. Gerber, ``Atomic Force Microscope,''
  \href{http://dx.doi.org/10.1103/PhysRevLett.56.930}{Phys. Rev. Lett. 56, 930 (1986)}.
\bibitem{fritz00}
  Fritz, J. {\sl et al.}, Translating biomolecular recognition into nanomechanics.
  \href{http://dx.doi.org/10.1126/science.288.5464.316} {{\sl Science} {\bf 288}, 316-318 (2000)}.
\bibitem{rugar04}
  Rugar, D., Budakian, R., Mamin, H. J. \& Chui, B. W., Single spin detection by magnetic resonance force microscopy.
  \href{http://dx.doi.org/10.1038/nature02658} {{\sl Nature} {\bf 430}, 329 (2004)}.
\bibitem{chaste12}
  Chaste, J. {\sl et al.}, A nanomechanical mass sensor with yoctogram resolution.
  \href{http://dx.doi.org/10.1038/NNANO.2012.42} {{\sl Nat. Nanotechnol.} {\bf 7}, 300-303 (2012)}.
\bibitem{hanay12}
  Hanay, M. S. {\sl et al.}, Single-protein nanomechanical mass spectrometry in real time.
  \href{http://dx.doi.org/10.1038/nnano.2012.119} {{\sl Nat. Nano.} {\bf 7}, 602-608 (2012)}.
\bibitem{poot12}
  Poot, M. \& Zant, H. S. J. Van der, Mechanical systems in the quantum regime.
  \href{http://dx.doi.org/10.1016/j.physrep.2011.12.004} {{\sl Phys. Rep.-Rev. Sec. Phys. Lett.} {\bf 511}, 273-335 (2012)}.

\bibitem{yang02}
  Yang, J., Takahito, O. \& Esashi, M., Energy Dissipation in Submicrometer Thick Single-Crystal Silicon Cantilevers.
  \href{http://dx.doi.org/10.1109/JMEMS.2002.805208} {{\sl J. Microelectromech. Syst.} {\bf 11}, 775-783 (2002)}.  
\bibitem{usenko11}
  Usenko, O., Vinante, A., Wijts, G. \& Oosterkamp, T. H., A superconducting quantum interference device based read-out of a subattonewton force sensor operating at millikelvin temperatures.
  \href{http://dx.doi.org/10.1063/1.3570628} {{\sl Appl. Phys. Lett.} {\bf 98}, 133105 (2011)}.
\bibitem{rast06}
  Rast, S. {\sl et al.}, Force microscopy experiments with ultrasensitive cantilevers.
  \href{http://dx.doi.org/10.1088/0957-4484/17/7/S15} {{\sl Nanotechnology} {\bf 17}, S189 (2006)}.
\bibitem{teufel09}
  Teufel, J. D., Donner, T., Castellanos-Beltran, M. A., Harlow, J. W. \& Lehnert, K. W., Nanomechanical motion measured with an imprecision below that at the standard quantum limit.
  \href{http://dx.doi.org/10.1038/nnano.2009.343} {{\sl Nat. Nano.} {\bf 4}, 820-823 (2009)}.


\bibitem{hutchinson04}
  Hutchinson, A. U. {\sl et al.}, Dissipation in nanocrystalline-diamond nanomechanical resonators.
  \href{http://dx.doi.org/10.1063/1.1646213} {{\sl Appl. Phys. Lett.} {\bf 84}, 972-974 (2004)}.


\bibitem{parikh92}
  Parikh, N. R. {\sl et al.}, Single-crystal diamond plate liftoff achieved by ion-implantation and subsequent annealing.
  \href{http://dx.doi.org/10.1063/1.107981} {{\sl Appl. Phys. Lett.} {\bf 61}, 3124-3126 (1992)}.
\bibitem{aharonovich12}
  Aharonovich, I. {\sl et al.}, Homoepitaxial growth of single crystal diamond membranes for quantum information processing.
  \href{http://dx.doi.org/10.1002/adma.201103932} {{\sl Adv. Mater.} {\bf 24}, OP54-OP59 (2012)}.
\bibitem{zalalutdinov11}
  Zalalutdinov, M. K. {\sl et al.}, Ultrathin single crystal diamond nanomechanical dome resonators.
  \href{http://dx.doi.org/10.1021/nl202326e} {{\sl Nano Lett.} {\bf 11}, 4304-4308 (2011)}.

\bibitem{supplementary}
  See supplementary information accompanying this manuscript.
\bibitem{lee08}
  Lee, C. L., Gu, E., Dawson, M. D., Friel, I. \& Scarsbrook, G. A., Etching and micro-optics fabrication in diamond using chlorine-based inductively-coupled plasma.
  \href{http://dx.doi.org/10.1016/j.diamond.2008.01.011} {{\sl Diam. Relat. Mat.} {\bf 17}, 1292-1296 (2008)}.






\bibitem{sque06}
  Sque, S. J., Jones, R. \& Briddon, P. R., Structure, electronics, and interaction of hydrogen and oxygen on diamond surfaces.
  \href{http://dx.doi.org/10.1103/PhysRevB.73.085313} {{\sl Phys. Rev. B} {\bf 73}, 085313 (2006)}.
\bibitem{panich10}
  Panich, A. M. {\sl et al.}, Structure and Bonding in Fluorinated Nanodiamond.
  \href{http://dx.doi.org/10.1021/jp9078629} {{\sl J. Phys. Chem. C} {\bf 114}, 774-774 (2010)}.
\bibitem{miller99}
  Miller, J. B., Amines and thiols on diamond surfaces.
  \href{http://dx.doi.org/10.1016/S0039-6028(99)00683-4} {{\sl Surface Sciences} {\bf 439}, 21-33 (1999)}.
  

\bibitem{degen09}
  Degen, C. L., Poggio, M., Mamin, H. J., Rettner, C. T. \& Rugar, D., Nanoscale magnetic resonance imaging.
  \href{http://dx.doi.org/10.1073/pnas.0812068106} {{\sl Proc. Nat. Acad. Sci. U.S.A.} {\bf 106}, 1313-1317 (2009)}.

\bibitem{watanabe99}
  Watanabe, H. {\sl et al.}, Homoepitaxial diamond film with an atomically flat surface over a large area.
  \href{http://dx.doi.org/10.1016/S0925-9635(99)00126-0} {{\sl Diamond and Related Materials,} {\bf 8}, 1272–1276 (1999)}.
\bibitem{davies92}
  Davies, G., Lawson, S. C., Collins, A. T., Mainwood, A. \& Sharp, S. J., Vacancy-related centers in diamond.
  \href{http://dx.doi.org/10.1103/PhysRevB.46.13157} {{\sl Phys. Rev. B} {\bf 46}, 13157-13170 (1992)}.
\bibitem{ishikawa12}
 Ishikawa, T. {\sl et al.}, Optical and Spin Coherence Properties of Nitrogen-Vacancy Centers Placed in a 100 nm Thick Isotopically Purified Diamond Layer.
  \href{http://dx.doi.org/10.1021/nl300350r} {{\sl Nano Lett.} {\bf 12}, 2083-2087 (2012)}.
  
\bibitem{burek12}
  Burek, M. J. {\sl et al.}, Free-Standing Mechanical and Photonic Nanostructures in Single-Crystal Diamond.
  \href{http://dx.doi.org/10.1021/nl302541e} {{\sl Nano. Lett.}, ASAP Article (2012)}.
\bibitem{maletinsky12}
 P. Maletinsky, {\sl et al.}, A robust scanning diamond sensor for nanoscale imaging with single nitrogen-vacancy centres
  \href{http://dx.doi.org/10.1038/NNANO.2012.50} {{\sl Nat. Nano.} {\bf 7}, 320-324 (2008)}.

\bibitem{ovartchaiyapong12}
  Ovartchaiyapong, P., Pascal, L. M. A., Myers, B. A., Lauria, P., \& Bleszynski Jayich, A. C., High quality factor single-crystal diamond mechanical resonators.
  \href{http://dx.doi.org/10.1063/1.4760274} {{\sl Appl. Phys. Lett.} {\bf 101}, 163505 (2012)}.

\end{thebibliography}
\end{document}